\documentclass[oldversion]{aa}
\usepackage{graphicx}        
\usepackage{psfig}        

\begin{document}

%
   \title{Fine structure in the phase space distribution of\\
    nearby subdwarfs }

   \author{M. I. Arifyanto and B. Fuchs}


   \institute{Astronomisches Rechen--Institut am Zentrum f\"ur Astronomie der
   Universit\"at Heidelberg, \\ M\"onchhofstrasse 12--14, 69120 Heidelberg,
   Germany}

   \date{Received 2005; accepted 2005}

   \abstract{We analysed the fine structure of the phase space distribution 
   function of nearby subdwarfs using data extracted from various catalogues.
   Applying a new search strategy based on Dekker's theory of galactic 
   orbits, we found four overdensely populated regions in phase space. 
   Three of them were correlated with previously known star streams: 
   the Hyades--Pleiades and Hercules streams in the thin disk of
   the Milky Way and the Arcturus stream in the thick disk. In addition we find 
   evidence for another stream in the thick disk, which resembles closely
   the Arcturus stream and probably has the same extragalactic origin.
      }
		
   \mail{fuchs@ari.uni-heidelberg.de}
   
   \titlerunning{Fine structure in the phase space distribution of nearby 
   subdwarfs}
    		
   \maketitle

%

\section{Introduction}

Fine structure in the velocity distribution of stars in the Milky Way was 
discovered and studied by O.J.~Eggen during almost all of his career
(Eggen 1996 and references therein). Some of Eggens's star streams are 
associated with young open clusters and can be naturally interpreted as clouds 
of former members, now unbound and drifting away from the clusters. Other
streams contain only stars older than 10 Gyrs.  
Since for many members distances were not known but had to be assumed in order
to construct space velocities, the real existence of such old streams has often 
been doubted.
 
However, modern data seem to confirm the concept of old star streams. 
Helmi et al.~(1999) found  the signature of a cold stream in the velocity 
distribution of the halo stars of the Milky Way when analyzing Hipparcos data. 
This was confirmed later by Chiba \& Beers (2000) using their own data (Beers
et al.~2000). Helmi et al.~(1999) interpreted this stream as part of the tidal 
debris of a disrupted satellite galaxy accreted by the Milky Way, which ended 
up in the halo. Indeed, numerical simulations have shown that relic stars from 
disrupted satellites can stay on orbits that are close together for many Gyrs 
(Helmi et al.~2003, Helmi 2004). These then show up as 
overdensities in phase space. In the same vein Navarro et al.~(2004)  
argue that Eggens's (1996) Arcturus group is another such debris stream, but 
in the thick disk of the Milky Way, dating back to an accretion event 5 to 8 
Gyrs ago. These observations complement observations of ongoing satellite
accretion such as of the Sagittarius dwarf galaxy (Ibata et al.~1994) or very
recent accretion in the form of the Monoceros stream discovered in the outer 
disk of the Milky Way with SDSS data (Newberg et al.~2002, Yanny et al.~2003, 
Rocha--Pinto et al.~2003, Pe\~{n}arrubia et al.~2005). Extended periods of the
accretion of satellites onto massive galaxies are also theoretically expected.
For instance, recent sophisticated simulations of the formation of a disk 
galaxy in the framework of cold dark matter cosmology and the cosmogony of 
galaxies by Abadi et al.~(2003a, b) suggest that disrupted satellites 
significantly contribute not only to the stellar halo but also to the disk of 
a galaxy.

Old moving groups are also observed in the velocity distribution of thin disk 
stars in the solar neighbourhood. Using {\sf Hipparcos} parallaxes and 
proper motions, Dehnen (1998) found  new evidence of the Sirius--UMa, 
Pleiades--Hyades, and Hercules star streams by statistical methods.
Even more convincingly these streams show up in the extensive data sample of
the three--dimensional kinematical data of 
F and G stars in the solar neigbourhood by Nordstr\"om et al.~(2004, hereafter
NMA+). The crowding of these stars on orbits in certain parts of 
velocity space is attributed to dynamical effects. Dehnen (2000) and Fux (2001)
demonstrate that the Hercules stream may well be due to an outer Lindblad 
resonance of the stars with the central bar of the Milky Way. The Sirius--UMa
and Pleiades--Hyades streams, on the other hand, are probably due to orbital 
resonances of stars in the solar neighbourhood with spiral density waves in the 
Milky Way disk (De Simone et al.~2004, Quillen \& Minchev 2005). However, there
are also hints that further overdensities in velocity space might be relics of
accreted satellites (Helmi et al.~2005).

Here we use our own data (Arifyanto et al.~2005; hereafter 
AFJW) on the kinematics of nearby subdwarfs and develop a new strategy to 
search for signatures of old star streams in the phase space distribution of 
the stars. We then cross check our findings with the NMA+ data.

\section{Data and search strategy for streams}

\subsection{Data}
 AFJW construct their data set from the sample of F and G subdwarfs of
Carney et al.~(1994, hereafter CLLA), which is based on the {\sf Lowell Proper
Motion Survey}, the so--called Giclas stars. While keeping the precise radial 
velocity and metallicity data of CLLA, AFJW have significantly improved the 
accuracy of the distances and proper motions for a subset of the CLLA sample. 
The original CLLA sample contains 1464 stars, but kinematical and 
metallicity data are not available for every star. Many of the CLLA 
stars were observed with {\sf Hipparcos}, and AFJW identified 483 stars in the 
astrometric {\sf TYC2+HIP} catalogue (Wielen et al.~2001) and replaced the  
parallaxes and proper motions of CLLA by Hipparcos parallaxes and proper 
motions, respectively. The Hipparcos parallaxes were then used to recalibrate
the photometric distance scale for the rest of the CLLA stars. AFJW could 
identify 259 CLLA stars in the {\sf Tycho--2} catalogue (H{\o}g et al.~2000)
and adopted the proper motions given there. Thus the sample of AFJW that 
forms the basis of our analysis contains 742 subdwarfs with greatly improved 
parallax and proper motion data. While the photometric distances were corrected
by a factor of about 10\%, the old {\sf NLTT} proper motions were improved from 
an accuracy of 20 to 30 mas/yr to 2.5 mas/yr.

\subsection{Search strategy}

The aim of our search is to find overdensities of stars in phase space on 
orbits that stay close together. For that purpose we use Dekker's (1976) 
theory of galactic orbits. Since the latter is not well
known despite its usefulness, we repeat the basic steps here to estimate 
the parameters of stellar orbits. The first step is to separate the planar
from the vertical motion of a star. This assumption is justified, because we
are treating orbits of stars with disklike kinematics. Concentrating now on 
the planar motion in the galactic plane, the equation of motion of a star 
moving in the meridional plane is given by 
\begin{equation}
\ddot{R}= -\frac{\partial \Phi_{\rm eff}}{\partial R} 
=-\frac{\partial}{\partial R} \left( \Phi(R) + \frac{1}{2}\frac{L^2}{R^2}
\right)\,,
\end{equation}
where $R$ denotes the galactocentric radius. The effective potential 
$\Phi_{\rm eff}$ is constructed in the usual way with both the gravitational 
potential $\Phi (R)$, which is assumed to be axisymmetric,
and the vertical component of the angular momentum of the star $L$ .
Dekker's theory proceeds then like standard epicycle theory by choosing a mean 
guiding centre radius for the orbit of a star $R_0$ by setting
\begin{equation} 
 L = R_0^2\Omega (R_0) \quad {\rm with} \quad 
 \Omega(R) = \sqrt{\frac{1}{R} \frac{\partial \Phi}{\partial R}}
\end{equation} 
the mean angular frequency of a stellar orbit. The energy of a star on the 
circular mean guiding centre orbit is obviously given by
\begin{equation}
E_0 = \Phi(R_0)+\frac{1}{2}R_0^2\Omega^2(R_0)\,.
  \end{equation}
Furthermore the epicyclic frequency $\kappa_0$ is introduced, 
$\kappa^2(R_0)= 4 \Omega^2(R_0)\big[ 1 +\frac{1}{2} d\ln{\Omega}/
d\ln{R}|_{\rm R_0} \big]$. The key point of Dekker's (1976) formalism is to
expand the potential with respect to $\frac{1}{R}$ around $\frac{1}{R_0}$ as 
\begin{eqnarray}
\Phi(R)&=&\Phi(R_0) + \frac{d \Phi}{d (\frac{1}{R})}\bigg|_{\rm R_0}
\left(\frac{1}{R}-\frac{1}{R_0}\right)\\ \nonumber &+& \frac{1}{2}
\frac{d^2 \Phi}{d (\frac{1}{R})^2}\bigg|_{\rm R_0}
\left(\frac{1}{R}-\frac{1}{R_0}\right)^2 \,,
\end{eqnarray} 
which is asymmetric with respect to $R_0$ and thus more realistic than the 
Taylor expansion of $\Phi(R)$ in the standard epicycle theory. Expression (4) 
is written as 
\begin{equation}
\Phi(R) = a_0-\frac{b_0}{R} +\frac{c}{R^2}
\end{equation}
with the coefficients $a_0 = E_0 +\frac{1}{2}R_0^2 \kappa_0^2$, 
$b_0 = R_0^3 \kappa_0^2$, and $c_0 = \frac{1}{2}R_0^4( \kappa_0^2-\Omega_0^2)$.
The turning points of the radial motion  of a star $R_{\rm t}$ are defined by 
the condition $E=\Phi _{\rm eff}(R_{\rm t})$. If the potential (5) is inserted, 
this leads to 
\begin{equation}
\frac{R_{\rm t}}{R_0} = \frac{1}{1\pm e} \quad {\rm with} \quad
e = \sqrt{ \frac{2 (E-E_0)}{R_0^2 \kappa_0^2} } \,.
\end{equation} 
The orbits are thus characterised by the two isolating integrals of motion 
angular momentum $L$ and energy $E$. By comparing her approximation (4) with 
various forms of exact potentials Dekker (1976) has shown that it gives 
reliable results up to eccentricities of $e\approx 0.5$. $L$ and $e$ can be 
estimated directly for each star in our sample. We assume that every star
is essentially at the position of the Sun and find 
\begin{equation}
L = R_{\odot}(V+V_{\rm LSR}) = R_0 V_{\rm LSR}\,.
\end{equation} 
Here $R_{\odot}$ denotes the galactocentric distance of the Sun, for 
which we adopt 8 kpc, $V$ is the velocity component of the star pointing 
into the direction of galactic rotation, and $V_{\rm LSR}$ is the circular 
velocity of the local standard of rest, for which we adopt 220 km/s. 
The eccentricity $e$ is given by 
\begin{equation}
e_{R_{0}} = \sqrt{\frac{U^2+\frac{\kappa_0^2}
{\Omega _0^2}V^2}{R_0^2\kappa_0^2}}\,,
\end{equation} 
with $U=-\dot{R}$ the radial velocity component of the star. In the following we
assume a flat rotation curve implying $\kappa_0^2/\Omega _0^2 =2$ and 
$R_0^2\kappa_0^2 = 2\,V_{\rm LSR}^2$. The search for overdensities in phase
space of stars on essentially the same orbits is carried out in practice 
in a space spanned up by $\sqrt{U^2+2\,V^2}$ and $V$. In addition we study 
the distribution of stars in our sample in ($|W|$, $V$) velocity space. 
Since the Sun is located very close to the galactic midplane, the absolute 
value of the vertical velocity component$|W|$ is a measure of the energy 
associated with the vertical motion of a star.

\section{Results and discussion}

We split our sample up into two subsets with metallicities of [Fe/H]
$>$ -0.6 and [Fe/H] $\leq$ -0.6, respectively. 

\subsection{Thin disk}
 
The stars in our sample with metallicities [Fe/H] $>$ -0.6 dex have kinematics
of the old thin disk of the Milky Way. Of course the metallicity cut is
somewhat arbitrary, because the thin and thick disk populations do not have a
bimodal metallicity distribution, but the transition is quite gradual. In 
Fig.~1 we show the distribution of 
309 stars, which have $|W|$ velocities $<$ 50 km/s, over $\sqrt{U^2+2\,V^2}$
versus $V$ and $|W|$ versus $V$, respectively. The space velocities have been 
reduced to the local standard of rest by adding the solar motion 
$(U, V, W)_{\odot} = (10.0, 5.2, 7.2)$ km/s 
(Dehnen \& Binney 1998) to the observed space velocities. Instead of scatter 
plots we show colour coded wavelet transforms of our data in Fig.~1. For this
purpose we used the two--dimensional Mexican--hat wavelet transform 
described by Skuljan et al.~(1999). After some experimentation we found that a
wavelet scale of 10 km/s showed the overdensities in the data samples in the
clearest way. The Hercules stream at $V \approx$ -40 km/s is
clearly visible as is, to a lesser degree, the Hyades--Pleiades stream at
$V \approx$ -15 km/s, and in both cases exactly where expected (Dehnen 2000, 
NMA+). Since these streams have been discussed widely in the
literature, we do not go into any further details. We 
present them mainly to demonstrate that by recovering previously known 
streams our method is well--suited to searching for cold star streams. 
\begin{figure}
\centering
\vbox{
{\includegraphics[scale=0.5]{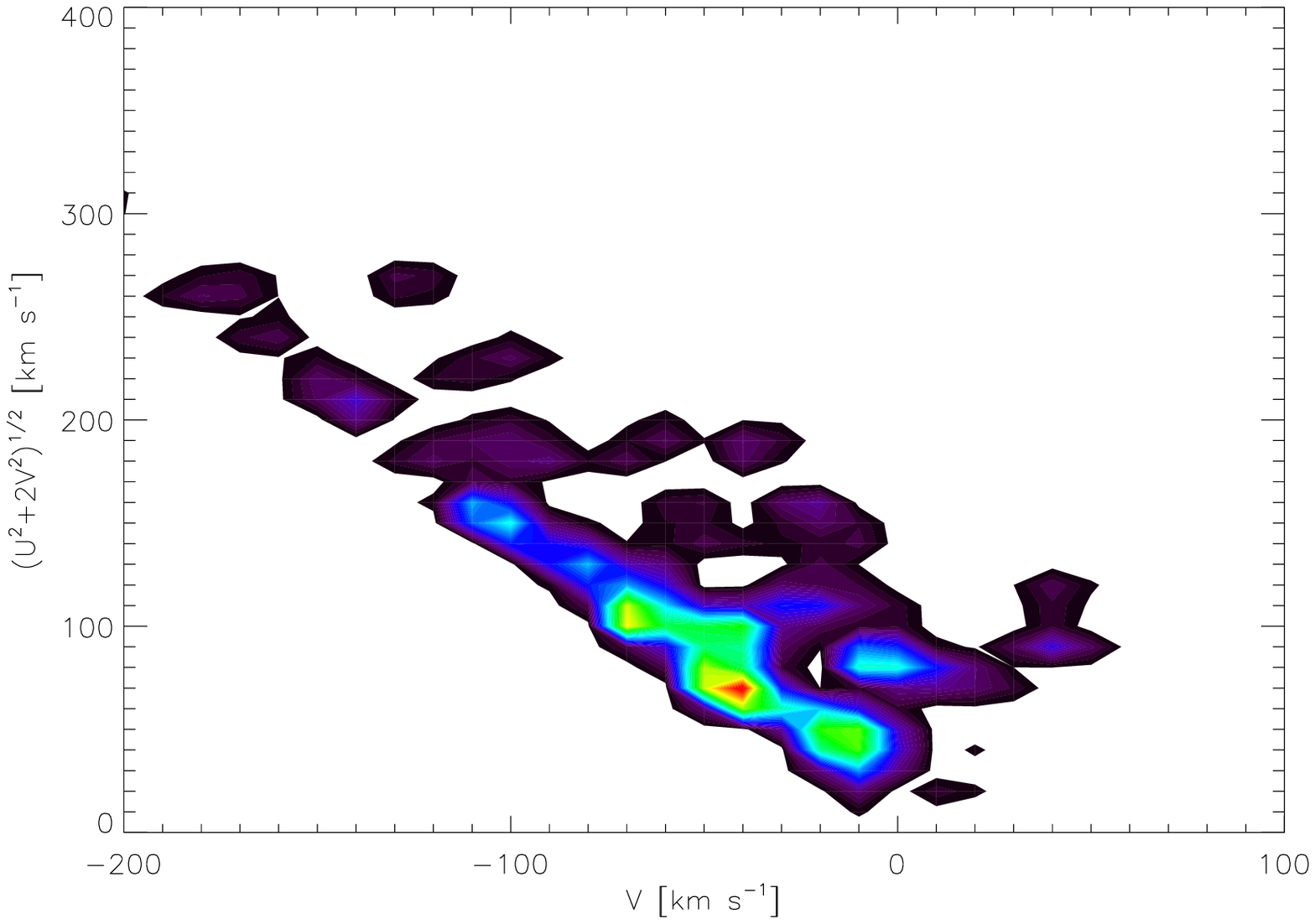}} \\  
{\includegraphics[scale=0.5]{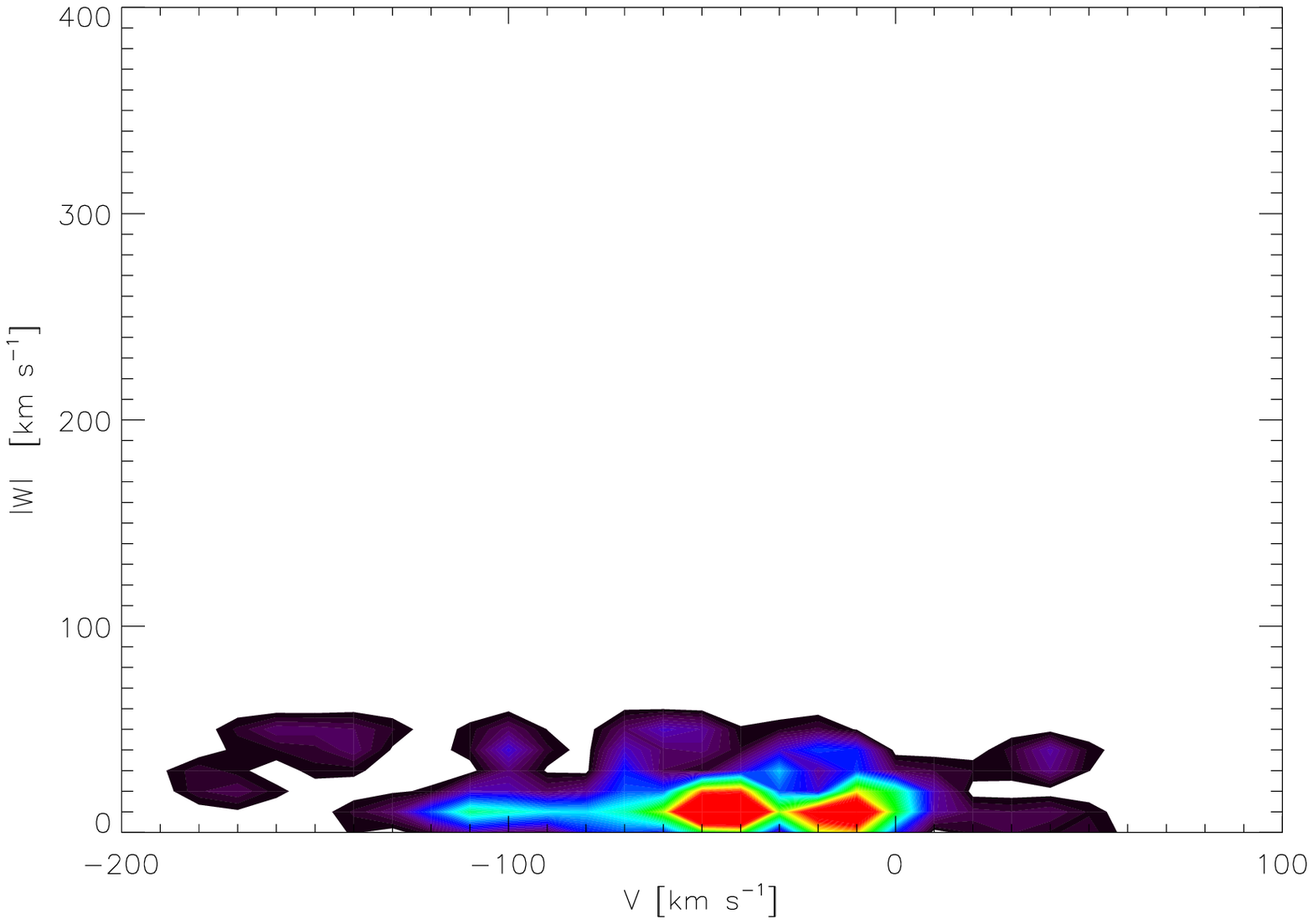}}}
\caption{Wavelet analysis of the distribution of thin disk stars over 
$\sqrt{U^2+2\,V^2}$ versus $V$ (top panel) and over $|W|$ versus $V$ 
(bottom panel). The wavelet scale of the Mexican hat kernel is 10 km/s and a
linear colour table from black over lilac, green, yellow to red is adopted.}
\label{fig:1}       
\end{figure}
On the other hand, it is instructive in order to assess the reality of such 
overdensities in phase 
space to compare the observed distribution with Monte Carlo simulations of 
realisations of a smooth distribution. In Fig.~2 we show such a Monte Carlo
simulation analysed in the same way as the observations. Three hundred nine 
stars were distributed
in the range -200 km/s $<$ $V$ $<$ 50 km/s and $\sqrt{U^2+2\,V^2}$ $<$ 300 km/s,
respectively, according to a Schwarzschild distribution
\begin{equation}
f \propto \exp{-\frac{1}{2}\left[ \left( \frac{U}{\sigma_{\rm U}}\right)^2
+ \left( \frac{V - \bar{V}}{\sigma_{\rm V}}\right)^2 \right] }\,,
\end{equation}
with parameters $\sigma_{\rm U}$ = 45 km/s, $\sigma_{\rm V}$ = 32 km/s,
and $\bar{V}$ = 26 km/s, which have been derived from the {\sf CNS4} catalogue 
as representative of disk stars in the solar neighbourhood 
(Jahrei{\ss} \& Wielen 1997). As can be seen from Fig.~2 the simulated
distributions looks generally smoother than the observed distribution but also 
show considerable Poisson fluctuations that can be confused
with real cold star streams. The only remedy for detecting real star streams is
obviously to search for such streams in separate data sets.
\begin{figure}
\centering
\vbox{
{\includegraphics[scale=0.5]{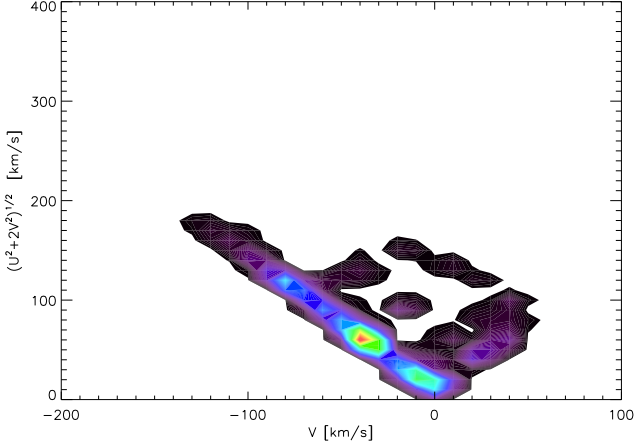}} }
\caption{Same as Fig.~1, but for 309 stars distributed randomly in
phase space according to a smooth Schwarzschild distribution}
\label{fig:2}       
\end{figure}

\subsection{Thick disk} 

The remaining stars of our sample with metallicities [Fe/H] $\leq$ -0.6 dex
belong to the thick disk and halo of the Milky Way. The distribution of 382
stars is shown in Fig.~3 in
the same way as above, but now restricted to $|W|$ $<$ 100 km/s.
There are two distinct features in the phase--space distribution function. 
The lesser feature at $V\approx$ -125 km/s corresponds to the familiar 
Arcturus stream (Eggen 1996, Navarro et al. 2004). The stars in this phase
space region are listed in Table 1 (available only in electronic form) 
giving all relevant data. The kinematics and metallicities of 
the stars listed in Table 1 can be compared with those of the stars considered 
by Navarro et al.~(2004; their Figs.~2 and 3) as members of the Arcturus group. 
Actually there is one common star, G2-34. The kinematics and metallicities
agree so well with each other that, even though the reality of low number 
overdensities is
difficult to assess, we are confident that both investigations have identified 
the same stream. Arcturus itself, although not a CLLA star, lies in 
Fig.~3 at $V$ = -114 km/s, $\sqrt{U^2+2\,V^2}$ = 165 km/s, and  $|W|$ = 4 km/s,
respectively. With a metallicity of [Fe/H] = -0.55 (Luck \& Heiter 2005), it 
fits well to the rest of the presumed stream members. We place the centre of
the stream at $V$ = -125 km/s and $\sqrt {U^2+2V^2}$ = 185 km/s implying 
$|U|$ = 55 km/s. According to Eq.~(7) the guiding centre radius of the 
orbits of the
stars now passing close to the Sun is $R_0 = 0.43\,R_{\odot}$ = 3.5 kpc. The
eccentricity is $e_{R_{0}}$ = 0.59 implying an outer turning radius of
$R_t = 2.5\,R_0$ = 8.5 kpc. The stars are apparently close to apogalacticon,
when they are at their slowest on their orbits and the detection probability 
is highest. In Fig.~4 we show a colour--magnitude diagram of the presumed
members of the Arcturus stream listed in Table 1. Overlaid are theoretical
isochrones of subdwarfs with an age of 12 Gyrs calculated for metallicities 
[Fe/H] = -0.5, -1, and -1.5, respectively (Yi et al.~2001). The good fit of the
isochrones indicates that the selected stars must be very old. In 
particular the tip of the main sequence fits well to the turn--off points of 
the isochrones. This is not a selection effect in the AFJW sample, because the
brightest stars in the sample have absolute magnitudes of $M_{\rm V}$ $<$ 4 
mag. Judging from the ages and metallicities of the stars and the 
similarity of their kinematics with that of debris from a disrupted satellite
in the vicinity of the Sun, we follow Navarro et al.~(2004) in concluding 
that the members of the Arcturus stream are of extragalactic origin.
\begin{figure}
\centering
\vbox{
{\includegraphics[scale=0.5]{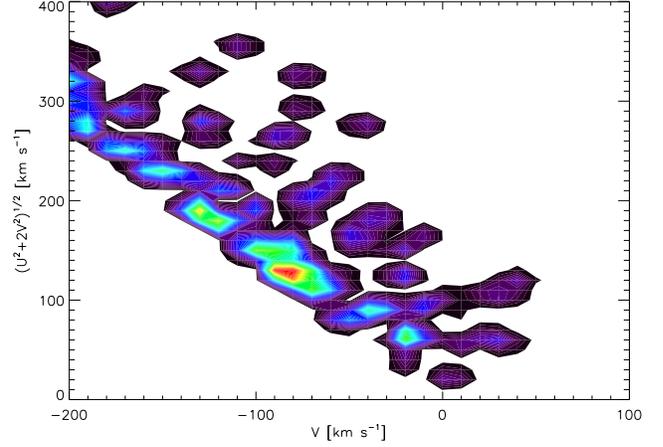}} \\ 
{\includegraphics[scale=0.5]{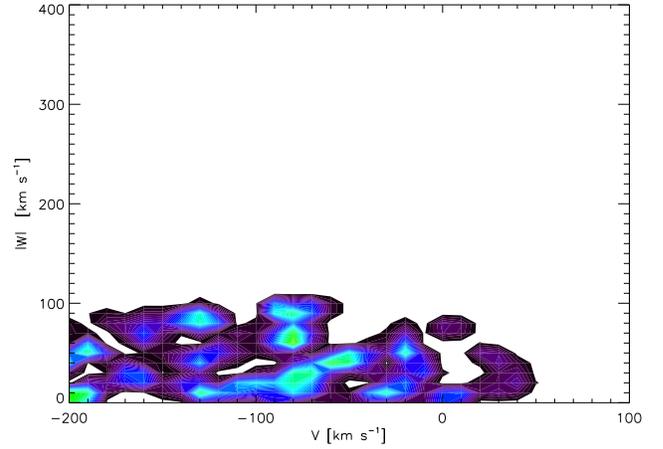}}}
\caption{Same as Fig.~1, but for thick disk stars.}
\label{fig:3}       
\end{figure}

\begin{figure}
{\includegraphics[scale=0.5]{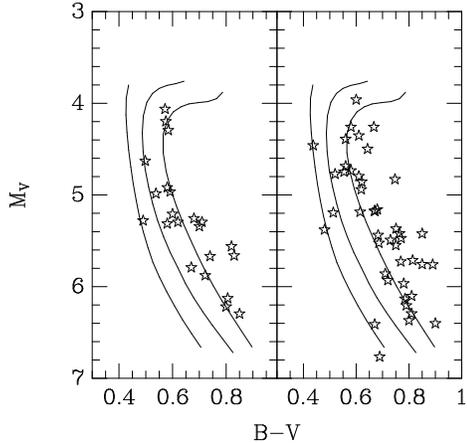}}
\caption{Colour--magnitude diagrams of the presumed members of the Arcturus
stream (left panel) and the proposed new stream (right panel). Overlaid are
theoretical isochrones for subdwarfs with an age of 12 Gyrs and and 
metallicities of [Fe/H] = -0.5, -1, and -1.5 (from right to left).}

\label{fig:4}       
\end{figure}

As can be seen in Fig.~3 there is a second strong feature in the phase--space
distribution of the thick--disk stars. This seems to be even more 
significant than the overdensity in the Arcturus region. The stars in this
overdensely populated region are listed in Table 2. To our knowledge the 
existence of a cold star stream in this part of phase space has not been 
suggested before. Comparing Figs.~3 and 1 we find a clear indication of a 
corresponding density enhancement at $V$ $\approx$ -70 km/s in Fig.~1.
This phase--space feature can thus also be traced among the more--metal rich 
stars, but is more prominently seen in the metal--poor population.

In order to test the robustness of our findings we analysed the
{\sf Copenhagen--Geneva Survey} of nearby F and G stars (NMA+). This is based
on {\sf Hipparcos} parallaxes, {\sf Tycho--2} proper motions, radial velocities,
and Str\"omgren photometry measured by the authors themselves. We have drawn
all those stars from the catalogue with metallicities [Fe/H] $<$ -0.6 and show 
a wavelet analysis of the phase--space distribution function of 591 stars in
Fig.~5 in the same way as in Figs.~1 and 3.
\begin{figure}
\centering
\vbox{
{\includegraphics[scale=0.5]{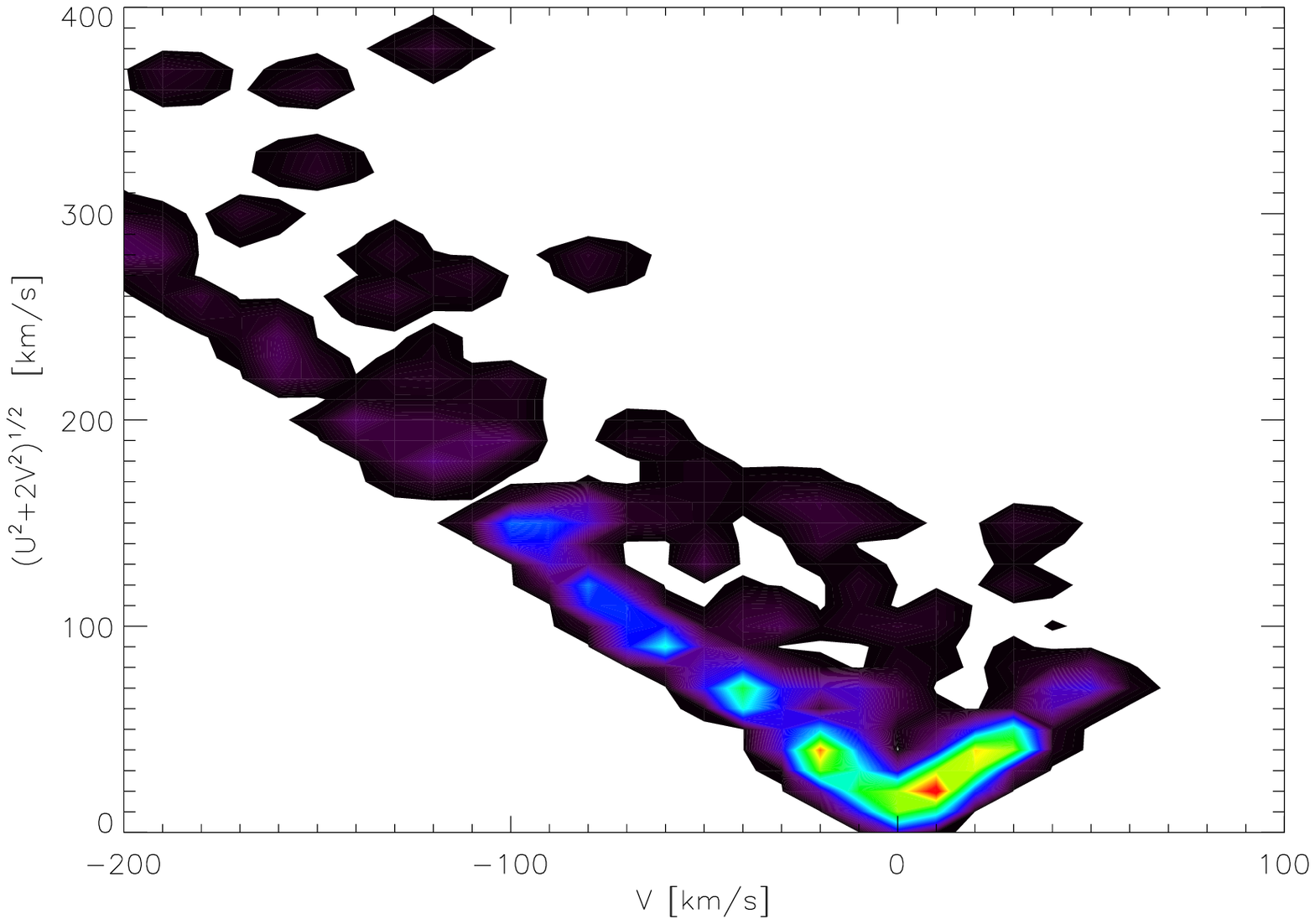}} \\ 
{\includegraphics[scale=0.5]{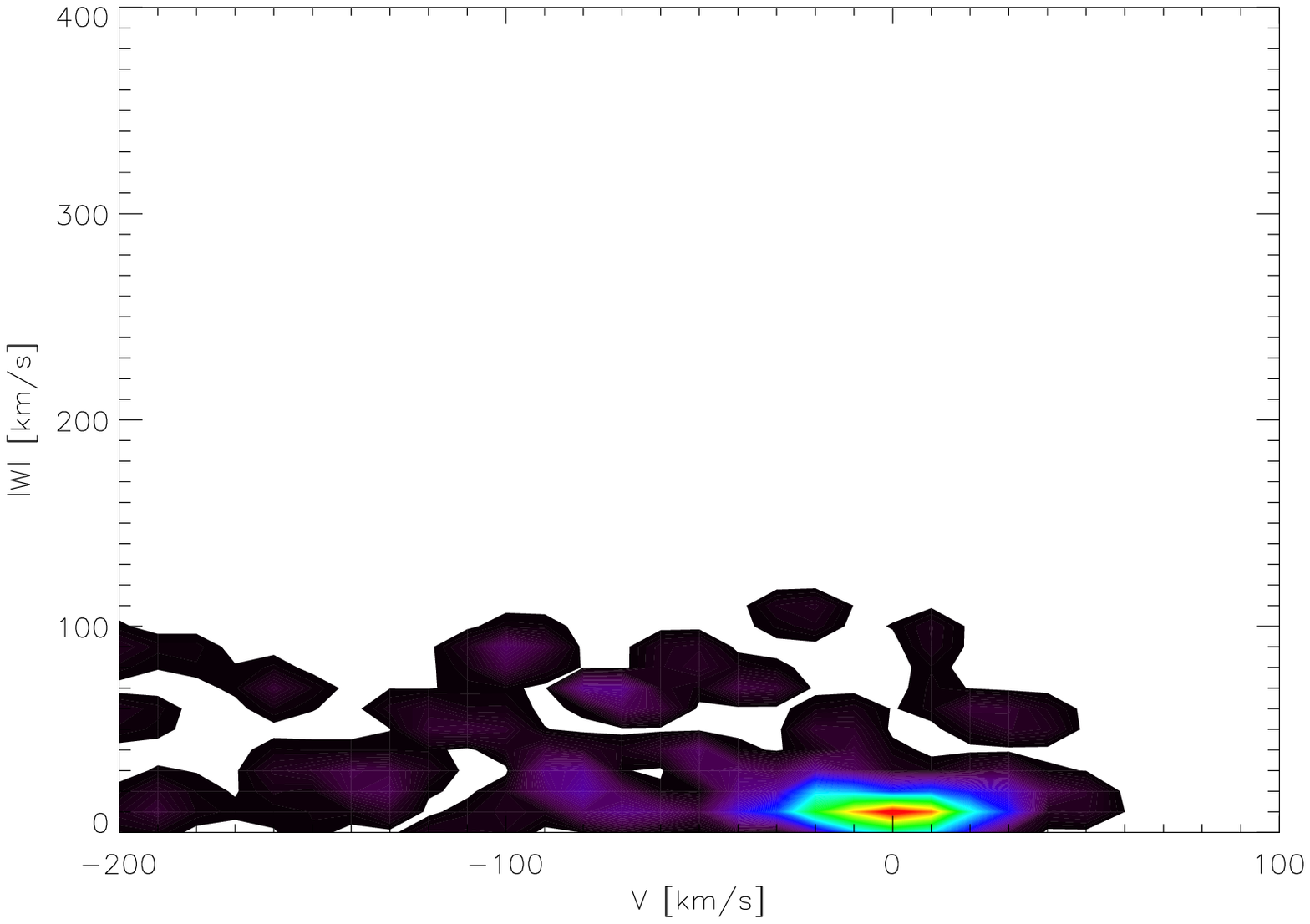}}}
\caption{Same as Fig.~1, but for stars  drawn from the Nordstr\"om et 
al.~(2004) sample with metallicities [Fe/H] $<$ -0.6.}
\label{fig:5}       
\end{figure}
Comparing Figs.~5 and 3 it becomes immediately clear that the NMA+
sample is much more fully populated at low space velocities. This reflects 
that the latter is kinematically unbiased, whereas the CLLA sample is 
biased towards high proper motion stars. Eggen's classical moving groups show 
up very clearly: the Sirius--UMa 
group at positive $V$ velocities, the Pleiades--Hyades stream at $V$ 
$\approx$ -20 km/s, and the Hercules stream at $V$ $\approx$ -40 km/s. 

Even 
though these moving groups were originally devised by Eggen among thin disk 
stars with metallicities of [Fe/H] $\approx$ 0, they are also very prominent
among the metal--poor stars. This indicates that the origins of the moving 
groups cannot be dissolving open clusters, but must be due to non--axisymmetric
perturbations of the gravitational potential of the Milky Way (Dehnen 2000,
Quillen \& Minchev 2005). In Fig.~5 there is a weak but significant sign of the
Arcturus stream, and some of the stars found as members of the Arcturus
stream in the AFJW sample appear in the NMA+ catalogue. In our view this is due
to the kinematical bias in the AFJW sample, so that the phase space is more 
richly populated at these negative $V$--velocities than in the NMA+ sample. For 
instance, in the NMA+ sample there are 144 stars with -200 km/s $<$ $V$ $<$ 
-50 km/s and $\sqrt{U^2+2\,V^2}$ $<$ 400 km/s, respectively, whereas in the 
AFJW sample there are 191 stars in the same range. However, the overdensity
between -100 km/s $<$ $V$ $<$ -60 km/s is clearly discernible in Fig.~5, which 
confirms the detection of the new cold star stream claimed above.

Our data as given in Tables 1 and 2 show that the velocity
and metallicity distributions of the members of the proposed new stream and the
Arcturus stream are practically identical. Also the colour--magnitude diagrams 
shown in Fig.~4 seem to indicate that the stars stem from the same population.
We place the centre of the proposed new stream at $V$ = -80 km/s and 
$\sqrt{U^2+2V^2}$ = 130 km/s implying $|U|$ = 64 km/s. The mean guiding centre 
radius of the orbits of these stars now passing close to the sun is 
$R_0=0.64\,R_{\odot}$ = 5.1 kpc. The eccentricity is $e_{R_{0}}$ = 0.42
and the outer turning radius is at $R_{\rm t} = 1.7\,R_0$ = 8.7 kpc. Thus 
the stars of the proposed new stream are also on their orbits close to 
apogalacticon.

Their orbits are actually
very similar to the orbits of the presumed members of the Arcturus stream.
We can at present only speculate about the possible origin of the stream. 
However, the similarity of the characteristics of the new stream with the 
Arcturus stream seems to point to an extragalactic origin. Moreover, both 
streams are probably related to each other. Indeed, Helmi et al.~(2005)  
show that in numerical simulations of the disruption of a satellite galaxy 
falling into its parent galaxy, the satellite debris can end up in several cold 
star streams with roughly the same characteristic eccentricities of their 
orbits. Precisely this seems to be the case here, so that both streams can have
very well originated from the same accretion event of a dwarf galaxy into the
Milky Way. How the star streams discussed here are related to the star streams 
reported by Helmi et al.~(2005), especially their groups 2 and 3, has yet to be
explored.

\acknowledgements{M.I.A.~acknowledges support for this work as part of a 
Ph.D.~thesis by a DAAD scholarship. We are grateful to the anonymous referee 
for his thoughtful comments. This research has made extensive use of the SIMBAD 
database, operated at CDS, Strasbourg, France.}

{}
\Online
\appendix  
\begin{table*}
\caption{Potential members of the Arcturus stream}
         \label{tabarc}
\[
\begin{tabular}{rrrrrrrrrr}
\hline
\hline
 \noalign{\smallskip}
  Name & RA[2000] & Dec[2000] & $V$ & Plx $\pi$ & $\mu^*_\alpha$ & $\mu_\delta$
   &  $\sigma_\pi$ & $\sigma_{\mu^*_\alpha}$ & $\sigma_{\mu_\delta} $ \\ 
Hip/Giclas & deg & deg & mag  & mas & mas/yr& mas/yr & mas/yr & mas/yr &
 mas/yr \\
 \noalign{\smallskip}
 \hline
 \noalign{\smallskip}
   13111 &    42.15594864  & 22.59843445  &  10.10 &   11.03  &    55.03 &  
   -359.47  & 1.55  & 1.07 &  0.97 \\
   36491 &   112.62090302  & 18.96128273  &   8.48 &   20.00  &    27.80 &  
   -436.75  & 1.45 &  0.95 &  0.62 \\
   36710 &   113.26815033  & 76.92041016 &   10.32  &  12.93  &   242.95 &  
   -201.59 &  1.42 &  0.96 &  1.28 \\
   40613  &  124.37228394 &  -3.98961496  &   7.74  &  20.46  &  -145.25 &  
   -438.59 &  1.12 &  0.88 &  0.91 \\
   53070 &   162.86718750 &  20.27749062  &   8.21  &  19.23  &  -260.72 &  
   -456.01 &  1.11 &  0.71 &  0.62 \\
   58253 &   179.20997620  & 13.37740707  &   9.95  &  11.98  &  -320.47 &  
   -173.65  & 1.49 &  0.90 &  0.73 \\
   74033 &   226.94375610 &   8.87977409  &   8.26  &  15.40  &  -518.44  &  
   -57.96  & 1.33 &  0.86 &  0.86 \\
   94931  &  289.75228882 &  41.63460541  &   8.87  &  28.28  &    98.78 &  
   -631.15 &  0.85 &  0.72 &  0.70 \\
  105888  &  321.67877197 &   5.44163942  &   8.49  &  13.02  &   167.04 &  
  -246.55 &  1.09 &  0.95 &  0.65 \\
  113514  &  344.83105469 &  12.19233322  &   8.35  &  20.59  &   332.17 &  
  -156.65 &  1.14 &  0.90 &  0.83 \\
   77637  &  237.74555969 &   8.42326450  &   9.95  &  10.07  &  -234.80 &  
   -159.80 &  1.39  & 1.47 &  1.51 \\
  G72-12  &   23.03840446 &  34.55577087  &  10.82  &   9.30  &   161.10 &  
  -223.20 &  1.37 &  1.80 &  1.80 \\
  G4-2    &   32.83348846 &   9.62153053  &  10.68  &   9.95  &   142.40 &  
  -267.70 &  1.47 &  2.50 &  2.30 \\
  G102-44 &   90.68071747 &  13.07698059  &  10.84  &   7.78  &   232.20 &  
  -148.30 &  1.15 &  1.50 &  1.50 \\
  G101-25 &   93.26070404 &  38.91038132  &  10.79  &   9.00  &   132.20 &  
  -199.60 &  1.33 &  3.60 &  3.40 \\
  G103-53 &  100.94363403 &  25.52507973  &  10.19  &  10.31  &    -7.60 &  
  -314.30  & 1.52  & 2.00 &  2.10 \\ 
  G42-34  &  150.80117798 &  19.84084129  &  10.70  &  13.16  &   -82.50 &  
  -343.60 &  1.95 &  1.50 &  1.60 \\
  G139-49 &  264.20046997 &   2.83879447  &  10.70  &  10.43  &  -137.80 &  
  -190.30 &  1.27 &  1.30 &  1.30 \\
  G204-30 &  267.49429321 &  37.52183914  &  10.27  &   9.72  &  -215.20 &  
  -165.50 &  1.44  & 3.20 &  3.00 \\
  G26-1   &  321.69931030 &  -8.39890003  &  11.27  &   6.34  &    56.70 &  
  -221.60 &  0.73  & 2.30 &  2.50 \\
  G265-43W &  325.39273071 &   85.91363525 &   10.52 &  13.80 &  239.80 &   
  108.50 &  2.04 &  2.80 &  3.00 \\
  G241-7  &  336.42810059 &  69.52659607  &  10.50  &   9.10  &  172.80 &    
  91.50 &  1.34 &  1.80 &  1.90 \\
 \noalign{\smallskip}
 \hline
 \hline
 Name & B-V & RV & $\sigma_{\rm RV}$ & [Fe/H] & U & V & W & & \\
 Hip/Giclas  & mag & km/s & km/s & dex & km/s & km/s & km/s & &  \\
 \noalign{\smallskip}
 \hline
 \noalign{\smallskip} 
 13111 & 0.580 & -22.3 &  0.7 &   -1.00 &  34.87 & -125.83 &  -88.74 & & \\
 36491 & 0.538 &  90.9 &   0.8 &   -0.81 &  -58.52 &  -124.68 &  -7.35 & & \\
 36710 & 0.722 &   -71.3 &  0.6 &   -0.60 &   33.97 & -122.70 &  47.65 & & \\
 40613 & 0.584 &   113.0 &  0.4 &   -0.51 &  -38.74 & -144.44 &  -43.19 & & \\
 53070  & 0.498 &    65.4 &   0.8 &   -1.56 &  -35.44 & -140.24 &   10.98 & &\\ 
 58253 & 0.700 &    28.9 &   0.6 &   -0.51 &  -80.23 & -122.25 &  -16.00 & & \\
 74033  & 0.575 &   -60.6 &   0.7 &   -0.89 & -114.20 & -125.63 &   25.19 & & \\
 94931 & 0.806 &  -121.1 &   0.6 &   -0.87 &   55.68 & -125.58 &  -85.21 & & \\
105888 & 0.572 &   -84.6 &   0.6 &   -0.80 &  -33.54 & -125.79 &  -44.34 & & \\
113514 & 0.580 &  -122.8 &   0.5 &   -0.67 &  -55.13 & -135.58 &   28.43 & & \\
77637 & 0.591 &   -51.6 &   1.0 &   -1.15 &  -37.11 & -138.26 &    6.81 & & \\
G72-12 & 0.830 &   -33.8 &   0.8 &   -0.25 &  -23.01 & -123.70 &  -70.70 & & \\
G4-2 & 0.740 & 38.3 & 0.3 & -0.80 & -20.31 & -122.47 & -83.21 & & \\
G102-44 & 0.710 & -28.8 & 0.5 & -0.62 & 61.06 & -136.59 & 81.41 & & \\
G105-25 & 0.820 & -47.6  &  0.4  &  -0.14  &  36.30  &-129.67  &   5.87  & & \\
G103-53 & 0.680 &     9.2 &   0.8 &   -0.70 &    0.39 & -130.49 &  -62.86 & & \\
G42-34  & 0.850 &    37.5 &   0.8 &   -0.81 &  -2.93 & -131.93 &  -13.93 & & \\
G139-49 & 0.670 &   -95.7 &   0.7 &   -1.23 &  -37.19 & -137.70 &  -14.74 & & \\
G204-30 & 0.600 &   -70.7 &   0.6 &   -0.98 &   48.82 & -136.30 &   39.61 & & \\
G26-1  & 0.490 &    14.5 &   0.9 &  -1.87 &   46.28 & -131.91 &  -99.60 & & \\
G265-43W & 0.800 &  -131.7 &   0.4 &  -0.76 &  -19.61 & -134.02 &  -84.70 & & \\
G241-7 & 0.620 &  -114.2 &   0.7 &  -0.97 &  -53.98 & -140.39 &  -28.18 & & \\
 \noalign{\smallskip}
 \hline
 \end{tabular}
 \]
\end{table*}
\begin{table*}
\caption{Potential members of the proposed new stream}
         \label{taberi}
\[
\begin{tabular}{rrrrrrrrrr}
\hline
\hline
 \noalign{\smallskip}
  Name & RA[2000] & Dec[2000] & $V$ & Plx $\pi$ & $\mu^*_\alpha$ & $\mu_\delta$
   &  $\sigma_\pi$ & $\sigma_{\mu^*_\alpha}$ & $\sigma_{\mu_\delta} $ \\ 
Hip/Giclas & deg & deg & mag  & mas & mas/yr& mas/yr & mas/yr & mas/yr & mas/yr \\
 \noalign{\smallskip}
 \hline
 \noalign{\smallskip}
    9080  &   29.23373795 &  11.66352558  &  10.52  &  13.26  &   378.47  &    
    2.28 &  1.97 &  1.38 &  1.38 \\ 				     
   10652 & 34.27974319 & 21.56681061 & 9.06 & 14.43 & 473.77 & 83.43 & 1.29 &
   0.77 & 0.69 \\
   11952 &   38.54603577 & -12.38429260  &   9.77  &   8.67  &    60.47  & 
   -185.07 &  1.78 &  1.21 &  1.24 \\ 				     
   16169  &   52.08785248 &  -6.53092098  &   8.23  &  21.98  &   358.02  & 
   -195.35 &  1.13  & 1.00 &  0.76 \\ 				     
   17147  &   55.09193802 &  -3.21697974  &   6.68  &  41.07  &   690.50  & 
   -213.58 &  0.85 &  0.86 &  0.79 \\ 				     
   22020  &   71.01499176 &  52.98161697  &   9.10  &  10.76  &    64.35  & 
   -294.97 &  1.38 &  1.04 &  1.00 \\ 				     
   22777  &   73.48664093 &  69.23905945  &   9.78  &  13.44  &   219.93  & 
   -124.84 &  1.54  & 0.92 &  1.11 \\  				     
   24030  &   77.48732758 &   5.55742788  &   9.71  &  10.29  &   269.99  &  
   -71.18 &  1.64  & 1.26 &  0.97 \\ 				     
   26452  &   84.41486359 &  68.73518372  &   9.60  &  13.14  &   245.55  & 
   -143.16 &   1.54 &  0.94 &  1.04 \\ 				     
   29814  &   94.17899323 &  47.06034470  &   9.18  &  20.39  &    57.31  & 
   -493.15 &  1.30 &  0.96  & 0.69 \\ 				     
   31740  &   99.60284424 &  48.79860687  &  10.11  &  11.92  &   131.60  & 
   -258.21 &  1.66 &  1.41 &  1.21 \\ 				     
   34642  &  107.62411499 &  53.25177765  &   8.80  &  10.77  &   -73.41  & 
   -241.43 &  1.21 &  1.02 &  0.83 \\ 				     
   50965  &  156.14868164 &  -5.51967478  &   9.80  &   9.70  &  -242.50  & 
   -166.16 &  1.40 &  1.05 &  1.00  \\				     
   61974  &  190.50057983 &  72.96403503  &   9.25  &  15.38  &  -287.50  &  
   -67.43 &  0.94 &  0.89 &  0.86 \\ 				     
   62607  &  192.43678284 &   1.18803751  &   8.13  &  30.12  &   -79.55  & 
   -644.49 &  0.91 &  0.61 &  0.48 \\ 				     
   73773  &  226.19566345 &  64.81214142  &   9.46  &  17.81  &  -255.17  &  
   -14.04 &  0.88 &  0.75 &  0.82 \\ 				     
   77122  &  236.21554565 &  62.86030579  &   8.95  &  11.53  &  -254.97  &  
   129.00  & 0.78 &  0.81 &  0.91 \\ 				     
   80700  &  247.14978027 &   3.25295258  &   8.81  &  21.50  &   -12.87  & 
   -526.94 &  1.27 &  0.93 &  0.88 \\ 				     
   89144  &  272.90850830 &  32.17737198  &  11.10  &   9.40  &   -96.72  & 
   -207.54 &  1.82  & 1.68 &  1.73 \\ 				     
   90365  &  276.59140015 &   8.61576462  &   8.32  &  26.30  &  -195.91  & 
   -468.58  & 1.05 &  0.88 &  0.75 \\  				     
   92918  &  283.97070312 &  -5.74521637  &   7.46  &  29.77  &  -200.21  & 
   -388.80 &  1.04  & 0.89 &  0.73 \\ 				     
  102923  &  312.77783203 &   7.02700377  &   9.82  &  20.71  &   237.77  & 
  -361.96 &  1.62 &  1.41 &  0.84 \\ 				     
  104913  &  318.77395630 &  62.84111404  &   9.56  &  14.51  &   122.87  &  
  260.66 &  0.89 &  0.87 &  0.89 \\ 				     
  112811  &  342.69140625 &   1.86516070  &   9.33  &  16.66  &   100.35  & 
  -384.38  & 1.33 &  0.97 &  0.84 \\ 				     
  114661  &  348.41174316 &  39.41738892  &  11.02  &  14.09  &   173.67  & 
  -313.88 &  2.18 &  1.63 &  1.28 \\ 				     
  115359  &  350.49285889 &  16.63253784  &   8.92  &  14.97  &   406.56  &  
  -49.04  & 1.22  & 0.93  & 0.86 \\ 				     
  118115  &  359.38964844 &  -9.64751911  &   7.89  &  20.98  &   454.84  & 
  -146.12 &  1.20 &  0.97 &  0.58 \\ 				     
  G30-46  &    2.05598330 &  15.00853062  &  11.01  &   8.91  &   220.70  &  
  -51.20 &  1.32  & 1.70  & 1.60 \\ 				     
  G69-21  &   11.66600418 &  33.82573700  &  10.34  &   9.20  &   250.20  &  
  -23.20 &  1.36 &  1.50  & 1.60 \\ 				     
  G71-33  &   26.30754089 &   3.51369452  &  10.63  &   8.91  &   224.10  &  
  -12.10  & 1.03 &  1.20 &  1.20 \\ 				     
  G5-44   &   53.57446671 &  22.98726463  &   9.18  &  10.83  &   150.70  & 
  -169.00  & 1.60 &  0.90 &  0.90 \\ 				     
  G78-41  &   53.73811722 &  38.30670166  &  10.21  &   9.83  &   144.10  & 
  -144.20 &  1.45 &  2.50 &  2.50 \\ 				     
  G99-40  &   88.23098755 &  -3.49025011  &   9.19  &  10.97  &   268.80  &  
  -49.80 &  1.62 &  1.20 &  1.10 \\ 				     
  G192-21 &   92.50205231 &  50.15151215  &   8.52  &  17.11  &   205.80  & 
  -265.70 &  2.53 &  1.20  & 1.30 \\ 				     
  G110-38 &  106.72556305 &  18.13643265  &  11.34  &   9.40  &     9.10  & 
  -167.70  & 1.39 &  1.70 &  1.70 \\ 				     
  G146-76 &  164.98948669 &  44.77882004  &  10.49  &  15.28  &  -101.80  & 
  -219.80  & 1.77 &  1.30 &  1.30 \\ 				     
  G10-12  &  169.80845642 &   5.67945290  &   9.29  &  25.16  &  -307.60  &  
  -74.20  & 3.72 &  1.40 &  1.40 \\ 				     
  G197-45 &  182.37043762 &  51.93362045  &  10.73  &  10.97  &  -235.30  & 
  -114.30 &  1.62 &  2.30  & 2.20 \\ 				     
  G66-51  &  225.20860291 &   2.12708616  &  10.63  &  11.11  &  -177.50  & 
  -109.30  & 1.36 &  1.40 &  1.40 \\ 				     
  G230-45 &  310.06958008 &  54.21994019  &  11.43  &   9.72  &    83.40  &  
  224.70  & 1.44  & 3.00 &  2.70 \\  				     
  G25-5   &  312.33590698 &   1.92505836  &  10.11  &  10.31  &   -43.60  & 
  -189.20  & 1.52 &  1.30 &  1.30 \\  				     
  G26-8   &  322.93942261 &  -1.92733061  &  10.47  &  11.41  &   203.00  &  
  -68.60 &  1.69 &  1.70 &  1.70 \\ 				     
  G28-16  &  341.91043091 &   6.42221117  &  11.59  &   7.99  &   250.60  &  
  -77.40  & 1.18 &  2.70 &  2.70 \\ 				     
  G67-40  &  345.44311523 &  11.82143307  &  10.66  &   9.50  &   286.70  &  
  -79.40 &  1.41 &  1.70 &  1.60 \\
 \noalign{\smallskip}
 \hline
 \end{tabular}
 \]
\end{table*}
\addtocounter{table}{-1}
\begin{table*}
\caption{cont.}
         \label{taberic}
\[
\begin{tabular}{rrrrrrrrrr} 				     
 \noalign{\smallskip}
 \hline
 \hline
 Name & B-V & RV & $\sigma_{\rm RV}$ & [Fe/H] & U & V & W & & \\
 Hip/Giclas  & mag & km/s & km/s & dex & km/s & km/s & km/s & &  \\
 \noalign{\smallskip}
 \hline
 \noalign{\smallskip} 
   9080  &  0.785 &   -10.7 &   0.9  &  -0.39 &  -93.51 &  -88.77 &  42.41  \\  
  10652  &  0.621 &   -21.1 &   0.9  &  -0.89 &  -99.83 &  -92.30 &  83.28  \\  
  11952  &  0.437 &    24.0 &   0.7  &  -1.82 &   27.76 &  -99.19 & -36.06  \\  
  16169  &  0.619 &    63.5 &   0.5  &  -0.58 &  -56.55 &  -90.59 & -19.11  \\  
  17147  &  0.554 &   120.3 &   0.6  &  -0.85 & -109.88 &  -85.71 & -44.86  \\  
  22020  &  0.667 &    30.2 &   0.3  &   0.20 &  -82.86 &  -89.87 & -60.51  \\  
  22777  &  0.850 &   -45.6 &   0.5  &  -0.42 &   -7.82 &  -97.26 &  22.69  \\  
  24030  &  0.520 &   -16.0 &   0.8  &  -0.92 &   10.42 &  -93.38 &  89.29  \\  
  26452  &  0.513 &   -35.6 &   0.7  &  -0.89 &   -7.89 & -100.11 &  41.20  \\  
  29814  &  0.769 &    22.6 &   0.4  &  -0.55 &  -54.64 &  -99.83 & -29.69  \\  
  31740  &  0.730 &    85.9 &   0.4  &  -0.61 & -101.93 &  -93.82 &  38.36  \\  
  34642  &  0.600 &   -28.6 &   0.6  &  -0.73 &  -23.64 &  -92.83 & -63.07  \\  
  50965  &  0.580 &    20.6 &   0.6  &  -0.63 &  -61.88 &  -93.89 & -91.75  \\  
  61974  &  0.615 &   -43.2 &   0.7  &  -0.86 &  -50.36 &  -85.29 & -18.44  \\  
  62607  &  0.686 &     2.4 &   1.0  &  -0.81 &   39.48 &  -84.29 & -42.29  \\  
  73773  &  0.814 &   -69.5 &   0.4  &   0.00 &  -24.27 &  -92.86 & -15.66  \\  
  77122  &  0.580 &   -48.8 &   7.2  &  -0.87 &  -88.72 &  -90.59 &  10.33  \\  
  80700  &  0.770 &    25.2 &   0.7  &  -0.17 &   80.96 &  -80.32 & -33.70  \\  
  89144  &  0.780 &   -38.0 &   0.8  &  -0.49 &   78.92 &  -92.41 &  -3.24  \\  
  90365  &  0.764 &   -18.1 &   0.4  &  -0.15 &   42.89 &  -82.43 &  -8.58  \\  
  92918  &  0.747 &   -73.4 &   0.3  &  -0.03 &  -31.25 &  -96.10 &   5.00  \\  
 102923  &  0.900 &   -61.8 &   0.6  &  -0.07 &  -21.53 &  -97.77 & -60.20  \\  
 104913  &  0.751 &   -64.6 &   0.6  &  -0.27 &  -74.40 &  -84.82 &  17.53  \\  
 112811  &  0.683 &    -4.1 &   0.8  &  -0.81 &   28.97 &  -89.88 & -62.26  \\  
 114661  &  0.689 &   -56.3 &   1.2  &  -2.68 &    9.71 &  -93.73 & -94.11  \\  
 115359  &  0.610 &   -40.4 &   0.5  &  -0.63 & -101.86 &  -85.55 & -27.46  \\  
 118115  &  0.643 &   -31.2 &   0.5  &  -0.02 &  -74.78 &  -83.84 &  -2.35  \\  
 G30-46  &  0.890 &   -22.6 &   0.6  &   0.20 &  -84.69 &  -86.10 & -21.34  \\  
 G69-21  &  0.680 &   -15.7 &   0.8  &  -0.57 &  -97.53 &  -86.43 &  -5.22  \\  
 G71-33  &  0.480 &    -9.6 &   1.5  &  -2.33 &  -82.74 &  -81.47 &  29.43  \\  
 G5-44   &  0.610 &    25.4 &   0.5  &  -0.06 &  -41.36 &  -90.59 & -23.50  \\  
 G78-41  &  0.670 &   -10.7 &   0.6  &  -0.65 &  -28.33 &  -94.13 & -10.79  \\  
 G99-40  &  0.560 &    49.6 &   0.6  &  -0.45 &  -27.56 &  -97.82 &  78.02  \\  
 G192-21 &  0.560 &   -18.6 &   0.6  &  -0.64 &   -3.60 &  -93.73 &  14.76  \\  
 G110-38 &  0.790 &    65.5 &   0.5  &  -0.67 &  -42.64 &  -96.55 & -17.96  \\  
 G146-76 &  0.670 &  -115.2 &   0.7  &  -2.31 &   41.50 &  -86.27 & -98.76  \\  
 G10-12  &  0.810 &   133.0 &   0.7  &  -0.92 &  -63.69 &  -96.28 &  88.98  \\  
 G197-45 &  0.720 &    23.4 &   0.6  &  -0.92 &  -77.58 &  -81.84 &  24.70  \\  
 G66-51  &  0.710 &  -118.8 &   0.4  &  -1.09 &  -96.02 &  -84.60 & -75.15  \\  
 G230-45 &  0.800 &   -79.8 &   0.7  &  -0.87 & -109.04 &  -87.34 &  22.67  \\  
 G25-5   &  0.670 &   -37.9 &   0.6  &  -0.66 &   34.72 &  -90.07 &  -9.38  \\  
 G26-8   &  0.850 &   -83.1 &   0.7  &  -0.32 &  -89.16 &  -80.25 & -21.01  \\  
 G28-16  &  0.810 &   -25.0 &   0.5  &  -0.86 & -105.96 &  -89.29 & -75.09  \\  
 G67-40  &  0.750 &   -29.3 &   0.4  &  -0.64 & -103.86 &  -90.36 & -62.80  \\  
 \noalign{\smallskip}
 \hline
 \end{tabular}
 \]
\end{table*}
\end{document}